\documentclass[12pt]{article}
\usepackage{epsfig}
\title{On noncommutative isotropic harmonic oscillator }
\author{Agnieszka Kijanka, 
Piotr Kosi\'nski\thanks{supported by KBN grant 5 P03B06021} 
\\Department of Theoretical Physics II \\
University of {\L}\'od\'z \\
Pomorska 149/153, 90 - 236 {\L}\'od\'z/Poland.}
\date{}

\begin{document}
\maketitle
\begin{abstract}
Energy spectrum of isotropic harmonic oscillator as a function of noncommutativity parameter $\Theta$\ is studied. It
is shown that for a dense set of values of $\Theta$\ the spectrum is degenerated and the algebra responsible for
 degeneracy can be always chosen
to be $sU(2)$. The generators of the algebra are constructed explicitely.
\end{abstract}

\newpage
It is well known  that the noncommutative geometry plays an important role in string theory and $M$-theory \cite{b1}. 
It has been found that, in a certain limit, string theory reduces to a gauge theory on noncommutative space. Since
then quantum theory on noncommutative spaces became a subject of intensive research.

The simplest example of such a theory is nonrelativistic quantum mechanics of fixed number of particles.
There are indications that some testable predictions can be found even in this case \cite{b2}. 
There appeared many papers dealing with quantum mechanics on noncommutative spaces \cite{b3} $\div$\ \cite{b20}. 

In particular, several authors studied the energy spectrum of natural hamiltonians on two dimensional noncommutative 
space defined by the commutation rule
\begin{eqnarray}
[\hat{x}_i,\;\hat{x}_j]=i\Theta\varepsilon_{ij},\;i,\;j=1,\;2\label{w1}
\end{eqnarray}
Unfortunately, most of these models cannot be solved exactly so only perturbative results are available.
 They show that the $\Theta$-dependence of physical quantities is quite involved. 

In particular, it has been suggested \cite{b21} \cite{b13} that, in general, the limit $\Theta \rightarrow 0$\
cannot be taken directly.

In the present note we show that, even if the limit $\Theta \rightarrow 0$\ \underline{can} be taken directly, still 
it is sometimes highly nontrivial (although numerically simple). This happens if 
the $\lim\limits_{\Theta \rightarrow 0}E_n(\Theta )$\ is not uniform in $n$.

More specifically, we consider isotropic harmonic oscillator on noncommutative plane (\ref{w1}). Both 
the hamiltonian and the commutation rule (\ref{w1}) are invariant under rotations. However, 
in two dimensions the rotation group is abelian so it does not imply energy degeneracy. 
In the commutative case the energy spectrum is degenerate due to the existence of dynamical $SU(2)$\ symmetry.
For $\Theta\not=0 $\ the situation 
appears to be more complicated. For a dense set of values of $\Theta$\ the energy spectrum \underline{is} 
degenerate and the dynamical $SU(2)$\ symmetry is still responsible for degeneracy; however, the structure of 
$SU(2)$\ multiplets changes abruptly with $\Theta$. The complement set, on which the hamiltonian has simple spectrum,
 is also dense. Therefore, with $\Theta$\ going to 0 one observes a complicated pattern of nondegenerate and
 degenerate states, the latter being organized in $SU(2)$\ multiplets varying with $\Theta$.

We start with the hamiltonian for isotropic oscillator on noncommutative plane
\begin{eqnarray}
\hat{H}=\frac{1}{2m}(\hat{p}^2_1+\hat{p}_2^2)+\frac{m\omega^2}{2}(\hat{x}_1^2+\hat{x}_2^2) \label{w2}
\end{eqnarray}

Let $x_i,\;p_i,\;i=1,\;2$\, be the standard canonical variables,
 $[x_i,\;x_j]=0,\;[p_i,\;p_j]=0,\;[x_i,\;p_j]=i\hbar \delta_{ij}$.
Then the commutation rules (\ref{w1}) can be solved in terms of canonical variables as follows
\begin{eqnarray}
&&\hat{x}_i=x_i-\frac{\Theta}{2\hbar}\varepsilon_{ij}p_j\label{w3} \\
&&\hat{p}_i=p_i \nonumber
\end{eqnarray}
With the above Ansatz eq.(\ref{w2}) takes the form 
\begin{eqnarray}
\hat{H}=\frac{1}{2M}(p_1^2+p_2^2)+\frac{M\Omega^2}{2}(x_1^2+x_2^2)-\frac{\Theta M \Omega^2}{2\hbar}L\label{w4}
\end{eqnarray}
where 
\begin{eqnarray}
L\equiv x_1p_2-x_2p_1 \label{w5}
\end{eqnarray}
is the angular momentum and
\begin{eqnarray}
&&M=\frac{m}{1+\frac{m^2\omega^2\Theta^2}{4\hbar^2}} \nonumber \\
&&\Omega=\omega\sqrt{1+\frac{m^2\omega^2\Theta^2}{4\hbar^2}}\label{w6}
\end{eqnarray}
It is straightforward to find the spectrum of $\hat{H}$. To this end we define the relevant creation and anihilation operators,
\begin{eqnarray}
a_{\pm}\equiv \frac{1}{2\sqrt{M\Omega\hbar}}(p_1\pm ip_2)-\frac{i}{2}\sqrt{\frac{M\Omega}{\hbar}}(x_1\pm ix_2);\label{w7}
\end{eqnarray}
then
\begin{eqnarray}
[a_{\alpha},\;a_{\beta}]=0,\;[a_{\alpha}^+,\;a_{\beta}^+]=0,\;[a_{\alpha},\;a_{\beta}^+]=\delta_{\alpha \beta}\label{w8}
\end{eqnarray}
and $\hat{H}$\ takes the form \cite{b8}
\begin{eqnarray}
\hat{H}=\hbar \Omega_+(N_++\frac{1}{2})+\hbar\Omega_-(N_-+\frac{1}{2})\label{w9}
\end{eqnarray}
$N_{\pm}$\ are the standard particle-number operators, $N_{\alpha}\equiv a_{\alpha}^+a_{\alpha}$, while
\begin{eqnarray}
\Omega_{\pm}\equiv \Omega\mp \frac{M\Omega^2\Theta}{2\hbar}\label{w10}
\end{eqnarray}
Note that, due to (\ref{w6}), $\Omega_{\pm}>0$. \\
The eigenvalues of $\hat{H}$\ read
\begin{eqnarray}
E_{n_+n_-}=\hbar\Omega_+(n_++\frac{1}{2})+\hbar\Omega_-(n_-+\frac{1}{2})\label{w11}
\end{eqnarray}
and the relevant eigenvectors are
\begin{eqnarray}
\mid n_+n_->=\frac{1}{\sqrt{n_+!}}\frac{1}{\sqrt{n_-!}}(a_+^+)^{n_+}(a_-^+)^{n_-}\mid 0>\label{w12}
\end{eqnarray}

The properties of spectrum depend on the ratio $\frac{\Omega_+}{\Omega_-}$. For irrational $\frac{\Omega_+}{\Omega_-}$\
the spectrum is nondegenerate while rational $\frac{\Omega_+}{\Omega_-}$\ leads to degeneracy. Consider the latter case. Assume that
\begin{eqnarray}
\frac{\Omega_+}{\Omega_-}=\frac{k}{l}\label{w13}
\end{eqnarray}
where $k,\;l$\ are relatively prime. Eq.(\ref{w13}) implies
\begin{eqnarray}
\Theta =\frac{2\hbar}{m\omega}\frac{\mid \frac{l-k}{l+k}\mid}{\sqrt{1-(\frac{l-k}{l+k})^2}}.\label{w14}
\end{eqnarray}
Let us put 
\begin{eqnarray}
\Omega_+=k\sigma, \;\Omega_-=l \sigma,\;\sigma  \equiv \frac{\omega}{\sqrt{lk}}\label{w15}
\end{eqnarray}
Then 
\begin{eqnarray}
E_{n_+n_-}=\hbar \sigma (kn_++ln_-)+\hbar \sigma (\frac{l+k}{2}) \label{w16}
\end{eqnarray}
The last term on $RHS$\ is an overall constant. It follows immediately from eq.(\ref{w16})
that the spectrum is degenerate, the level of degeneracy being equal to the number of natural solutions $n_{\pm}$\
to the equation $kn_++ln_-=const$. Suprisingly enough, the symmetry algebra responsible for degeneracy is
 always $sU(2)$, like for isotropic
case. To see this we classify all pairs $(n_+,\;n_-)$\ according to their congruence properties \cite{b22},
\begin{eqnarray}
&&n_+=pl+r_+ \;,\;\;\;0\leq \;r_+\leq\; l-1 \nonumber \\
&&n_-=qk+r_- \;,\;\;\;0\leq \;r_-\leq\; k-1 \label{w17}
\end{eqnarray}
Let us fix $r\equiv (r_+,\;r_-)$\ and let $X_r$\ be the subspace spanned by the vectors $\mid n_+\;n_->$\ such
that $n_+,\;n_-$\ are congruent to $r_+,\;r_-$ modulo $l$\ and $k$, respectively. Using (\ref{w17}) one can write the
energy spectrum of $\hat{H}$, when restricted to $X_r$, in the form
\begin{eqnarray}
E_{pq}=\hbar kl\sigma (p+q)+\hbar\sigma \left(k(r_++\frac{1}{2})+l(r_-+\frac{1}{2})\right)\label{w18}
\end{eqnarray}
Therefore, in each $X_r$\ the energy spectrum coincides, up to an additive constant, with that of isotropic oscillator. 
Moreover, the degeneracy
is possible only among the states belonging to the same $X_r$\ \cite{b22}.

It is also not difficult to find the relevant $sU(2)$\ symmetry algebra responsible for degeneracy in each $X_r$. To this
end let us note that each $X_r$\ is the Fock space if one makes an identification $\mid n_+\;n_->\cong \mid p,\;q>$\ 
and defines
\begin{eqnarray}
&&b_{r_+}\mid p,\;q>=\sqrt{p}\mid p-1,\;q>,\;b_{r_-}\mid p,\;q>=\sqrt{q}\mid p,\;q-1>\nonumber \\
&&b_{r_+}^+\mid p,\;q>=\sqrt{p+1}\mid p+1,\;q>,\;b_{r_-}^+\mid p,\;q>=\sqrt{q+1}\mid p,\;q+1>\label{w19}
\end{eqnarray}
New operators are obviously expressible in terms of $a_{\alpha},\;a_{\alpha}^+$. The relevant formulas 
are slightly complicated
and read \cite{b22}
\begin{eqnarray}
&&b_{r_+}=\sqcap_{s=1}^l(N_++s)^{-\frac{1}{2}}(\frac{N_+-r_+}{l})^{\frac{1}{2}}a_+^l \nonumber \\
&&b_{r_-}=\sqcap_{s=1}^k(N_-+s)^{-\frac{1}{2}}(\frac{N_--r_-}{k})^{\frac{1}{2}}a_-^k\label{w20}
\end{eqnarray}
In spite of their appearance these operators are well-defined.

Having constructed new creation-anihilation operators one easily finds symmetry operators. 
In fact, defining (here $\sigma_{i}$\ are Pauli
matrices)
\begin{eqnarray}
T_{ri}\equiv \frac{1}{2}b_{r\alpha}^+(\sigma_i)_{\alpha \beta}b_{r\beta},\;i=1,\;2,\;3\label{w21}
\end{eqnarray}
one checks the following relations hold in $\chi_r$:
\begin{eqnarray}
&&[T_{ri},\;\hat{H}]=0\label{w22}\\
&&[T_{ri},\;T_{rj}]=i\varepsilon_{ijk}T_{rk}\nonumber 
\end{eqnarray}
Now, the total Hilbert space is the orthogonal sum of the $X_r, \;X =\oplus_rX_r $.
 Let $P_r$\ be the projection operator on $X_r$; define
\begin{eqnarray}
T_i=\sum_rP_rT_{ri}P_r\equiv \sum_rT_{ri}P_r.\label{w23}
\end{eqnarray}
Then eqs.(\ref{w22}) extend to the whole Hilbert space $\chi$,
\begin{eqnarray}
&&[T_i,\;\hat{H}]=0\label{w24} \\
&&[T_i,\;T_j]=i\varepsilon_{ijk}T_k\nonumber
\end{eqnarray}
which proves that the symmetry algebra is always $sU(2)$\ provided eq.(\ref{w14}) holds.

It remains to construct explicitly $P_r$. Again, it is not difficult to verify that \cite{b23}
\begin{eqnarray}
P_r=\left(\frac{1}{l}\sum_{s=0}^{l-1}e^{\frac{2i\Pi s}{l}(N_+-r_+)}\right)\left(\frac{1}{k}\sum_{t=0}^{k-1}e^{\frac{2i\Pi t}{k}(N_--r_-)}\right)\label{w25}
\end{eqnarray}

Let us summarize our results. For $\Theta \rightarrow 0$\ all energy levels tend to their undeformed values.
 However, this limit is not uniform in quantum numbers $n_+,\;n_-$. This results in 
quite involved $\Theta$-behaviour of the system.
 For dense set of values the energy spectrum is nondegenerate and the hamiltonian is essentially
the only independent operator in the sense that any operator commuting with $\hat{H}$\ is a function of $\hat{H}$.
On the other hand, for the complement dense set of $\Theta$'s obeying eq.(\ref{w14}) the energy spectrum is
degenerate and the symmetry algebra responsible for this degeneracy is always $sU(2)$.
However, the structure of symmetry operators and irreducible $sU(2)$\ multiples change
 very rapidly with the change of $\Theta$.
In fact, a slight change of $\Theta$\ can produce an enormous change of $k,\;l$\ which determine
 the decomposition of $X$\
into the sum of $X_r$'s and the form of symmetry algebra.

This phenomenom has its classical counterpart. If the sympletic structure is modified by imposing 
$\{x_i,\;x_j\}=\Theta\varepsilon_{ij}$, the hamiltonian continues to be integrable for all values of $\Theta$.
However, it becomes superintegrable for a dense set of values of $\Theta$\ while it is not superintegrable 
for complement dense set.

\end{document}